\documentclass[twocolumn,showpacs,preprintnumbers,amsmath,amssymb]{revtex4}

\usepackage{graphicx}
\usepackage{dcolumn}
\usepackage{bm}

\newcommand{\noi}{\noindent}
\newcommand{\eq}{\begin{equation}}
\newcommand{\en}{\end{equation}}
\newcommand{\eqa}{\begin{eqnarray}}
\newcommand{\ena}{\end{eqnarray}}
\def\myre{{\rm Re}}

\newcommand{\real}{\relax{\rm I\kern-.18em R}}
\newcommand{\integer}{\relax{\rm I\kern-.18em N}}

\begin{document}
\title{Recombination of dyons into calorons \\
in $SU(2)$ lattice fields at low temperatures}
\author{E.-M. Ilgenfritz}
\affiliation{Humboldt-Universit\"at zu Berlin, Institut f\"ur Physik, 
Newtonstr. 15, D-12489 Berlin, Germany}
\author{B. V. Martemyanov}
\affiliation{ Institute for Theoretical and Experimental Physics, 
B. Cheremushkinskaya 25, 117259 Moscow, Russia}
\author{M. M\"uller-Preussker}
\affiliation{Humboldt-Universit\"at zu Berlin, Institut f\"ur Physik, 
Newtonstr. 15, D-12489 Berlin, Germany}
\author{A. I. Veselov}
\affiliation{ Institute for Theoretical and Experimental Physics, 
B. Cheremushkinskaya 25, 117259 Moscow, Russia}

\date{}

\begin{abstract}
By cooling of equilibrium lattice fields at finite temperature in $SU(2)$ 
gauge theory it has been shown that topological objects (calorons)
observed on the lattice in the confined phase possess a dyonic 
substructure which becomes visible under certain circumstances. 
Here we show that with the increasing temporal lattice extent
the distribution in the caloron parameter space is modified such that the
calorons appear non-dissociated into constituent dyons.
Still the calorons have nontrivial holonomy which is demonstrated by the
Polyakov line behaviour for these configurations. At vanishing temperature 
(on a symmetric lattice) topological lumps obtained by cooling show rotational 
symmetry in $4D$ for the action density, but a characteristic internal double 
peak structure of Polyakov lines with respect to all (temporal and spatial) 
directions.
\end{abstract}


\maketitle

\vspace{-5mm}
\section{Introduction}
\label{sec:introduction}
At high temperatures, near but below the deconfinement temperature, classical
solutions of Yang-Mills equations with nontrivial holonomy (KvB calorons
\cite{KvB,big_paper}) are seen on the lattice for $SU(2)$ gauge theory 
to be frequently dissociated into dyons~\cite{IMPV,IMMPV1,IMMPV2,IMMPSV}. 
This means that the distance between the dyons forming a caloron,
\eq
                d=\frac{\pi \rho^2}{b} \; ,
\label{eq:drhob}
 \en
is larger than the size of dyons 
(which is $b/\pi$ for the case of maximally nontrivial 
holonomy). Here $\rho$ is the instanton size parameter and $b$ is the temporal 
periodicity interval. 

These observations have been made by the use of cooling. Therefore, the question 
arises why such a substructure 
has not been observed by previous authors who have 
used the cooling method. In this paper we will argue that there is only a certain 
window of temperature or space-time asymmetry
where it can be revealed by this method. 

The interest in the existence of caloron constituents has increased since it has
been demonstrated in Ref.~\cite{GATTR-SCHAEFER} that a constituent substructure
very reminiscent of the caloron solutions can also be identified without cooling,
above and below the phase transition. This can be achieved by using the localization
properties of the fermionic zero modes of a suitably chirally-improved Dirac operator. 
The similarity with the properties of a caloron solution is strikingly realized for 
a certain fraction of configurations with topological charge $Q=\pm 1$,
where the single zero mode is seen to change
its localization when the periodicity of fermionic boundary conditions becomes
modified. A systematic study~\cite{GATTR-SCHAEFER} of the typical pattern of
localization and delocalization followed by jumps of the zero mode has revealed
that this pattern depends on the timelike holonomy exactly in a caloron-like way.
Whereas the topological density has a much more complicated structure, the positions
where the zero mode is pinned-down actually show the signatures expected for caloron
constituents~\cite{Lattice03}: 
they are local maxima of the topological density
$q(x)$ with a sign as required by the chirality of the mode. This suggests that
(dissociated or non-dissociated) calorons might really form the semiclassical
background of the gauge field near the phase transition.

Coming back to $SU(2)$ calorons with their two constituents, it seems that
the quantities $d$ and $\rho$ appearing in eq. (\ref{eq:drhob}) 
are impossible to be measured simultaneously: when $d$ is seen by observation 
of separate dyon positions no instanton-like profile (of topological density)
is observed which could be used to define $\rho$. When $d$ goes to zero ($d << b$) 
an instanton size parameter $\rho$ {\it can be measured} by comparison with the 
instanton's action density profile, but then {\it no dyons are seen} as separate 
objects. More precisely: the parameter $d$ cannot be measured for {\it all} caloron
configurations as the distance between constituents as long as only the action 
or topological charge densities are available as local observables to describe
them. 

The time periodicity parameter $b$ defines the temperature $T$: $b=1/T$. 
In order to demonstrate how the recombination of constituents depends on the  
temperature we can 
change the temporal extent $b$ of the lattice.

We employ the standard relaxation cooling technique using the Wilson lattice
action and concentrate on the investigation of lowest-action field configurations
with $~Q = \pm 1.$
Besides of the fact that solutions with $~|Q|=1~$ are not existing
on a torus in a mathematically strict sense \cite{BraamBaal},
also the cooling method has a limited relevance.
It cannot be used for revealing the full topological structure of the QCD
vacuum even if the latter is semi-classical to a certain extent. In particular,
it considerably weakens the chiral condensate compared with its value for
equilibrium fields \cite{Wol}.
Moreover, it is well-known that the Wilson action depends on the instanton
(or caloron) scale-size. Therefore, its minimization shrinks the localized
solutions until they disappear 'through the lattice meshs'. Thus, small
excitations will be lost first. In the literature - besides fermionic methods -
there are better techniques like
improved cooling, smoothing, smearing etc. allowing to decipher the topological
long-range structure of the gauge fields. But this task is {\bf not} our concern in
this paper. Here we ask the more modest question, what kinds of simple classical
solutions can be found from equilibrium gauge fields at different temperatures 
by successively minimizing the action.

This question remains interesting for those who want to build or apply semi-classical-like
approximations, i.e  models like the instanton gas or liquid model \cite{inst_gas} 
being successful in many phenomenological applications for which chiral symmetry
breaking (and not quark confinement) is playing the major r\^ole 
\cite{inst_reviews}. Until now 
most of the cooling (or smoothing) results obtained at $~T < T_c~$ have been 
interpreted in terms of BPST instanton \cite{BPST} and Harrington-Shepard (HS) 
caloron \cite{HS} solutions, respectively, all exhibiting trivial asymptotic
holonomy. 
Here we would like to convince the reader that the simplest caloron or instanton
solutions seen after cooling have typically non-trivial holonomy, irrespectively of
their possible dissociation into dyons pairs. Therefore, they cannot be correctly 
interpreted as BPST or HS solutions.     

The paper is organized as follows.
In section \ref{sec:nonstaticity} we draw the attention to the static nature of
configurations near the deconfining transition. Section \ref{sec:recombination}
presents our results on semiclassical configurations at finite temperatures,
pointing out the loss of ''staticity'' and the increasing importance of the
Polyakov loop for detecting the non-trivial substructure at lower temperature.
In section \ref{sec:torus} we extend the cooling studies to the symmetric torus.
We emphasize 
that the KvB caloron soultions were constructed in an infinite spatial
volume. When used at finite volume considerable deviations are to be expected 
when the spatial box size is no longer large compared to $b$, since the typical 
size of the constituents is $b/\pi$.
Finally, in section \ref{sec:conclusions}, we discuss the consequences, also in
the perspective of a twin paper by Gattringer {\it et al.}.

\section{Non-staticity and separation into constituents}
\label{sec:nonstaticity}
It turns out that the possibility to observe the dyonic constituents of
a KvB caloron as lumps of action depends on $(\frac{\pi \rho}{b})^2$.
In $SU(2)$ LGT at $\beta \equiv 4/g_0^2 = 2.2$ on a lattice $16^3\times 4$
the parameter $\rho$ is concentrated near the value $\rho \approx 2.5 a$ 
($a$ is the lattice spacing)~\cite{IMMPV1}. With $b=4 a$, 
$$(\frac{\pi \rho}{b})^2\approx 4~ >> ~1 \; .$$ 
This means that dyons are well separated. The value of $\rho$ in \cite{IMMPV1} 
was determined by fitting the lattice caloron with the analytic KvB caloron,
and formula (\ref{eq:drhob}) was used. On the lattice $16^3\times 6$ (with $b=6 a$) 
and at the same $\beta = 2.2$ ({\it i.e.} at a temperature $1.5$ times lower) 
the parameter $(\frac{\pi \rho}{b})^2$ would be of the order $O(1)$. Then, from
this simple arithmetics, one would expect that calorons are not dissociated into
dyonic lumps anymore.

The possibility to measure the distance between dyons inside a caloron just by
detecting the peaks of the action density on the lattice is given only in the case
of well-separated objects. Indirectly this distance can be measured by measuring
a quantity that can be called non-staticity~\footnote{See Ref.~\cite{IMMPSV} where
it has been defined in a slightly different way.}. 
Unfortunately, cooling yields metastable plateaus only for 
temperatures below the deconfining temperature. On the other hand, this allows us 
to restrict ourselves in the following to {\it maximally non-trivial holonomy} 
because the average Polyakov line vanishes. To describe for this special case
the relation between 
distance $d$ and non-staticity, we have considered analytical caloron solutions in
continuous space-time. 
We divided the time interval $b$ into $N_t$ time slices 
and expressed the action in the $i$-th time slice, $S_i=\sum_{\vec x} s_{{\vec x},i}$, 
in terms of the local action density $s_{{\vec x},i}$. The non-staticity $\delta_t$
is defined as
\eq
\delta_t = 
\frac{ \sum_{i=1}^{N_t}~\sum_{\vec x}
            |s_{{\vec x},i+1}-s_{{\vec x},i}|   }
     { \sum_{i=1}^{N_t}~\sum_{\vec x} 
             s_{{\vec x},i}          }  \; . 
\label{eq:nonst1}
\en
Obviously, this definition depends on the number of time slices $N_t$ such that
one would have $\delta_t \rightarrow 0$ for $N_t\rightarrow \infty$. In order to 
get an asymptotically $N_t$ independent quantity we will modify the definition of
$\delta_t$ as follows
\eq
\delta_t = 
\frac{ \sum_{i=1}^{N_t}~\sum_{\vec x}
            |s_{{\vec x},i+1}-s_{{\vec x},i}|  }
     { \sum_{i=1}^{N_t}~\sum_{\vec x} 
             s_{{\vec x},i}                    } \frac{N_t}{4} \; .
\label{eq:nonst2}
\en
The factor $1/4$ has been chosen such that for a lattice $16^3\times 4$, where 
all the simulations were initially done using (\ref{eq:nonst1}), the two definitions 
(\ref{eq:nonst2}) and (\ref{eq:nonst1}) agree. 

With the definition (\ref{eq:nonst2}) at hands we can calculate the non-staticity
of an analytic KvB caloron. The non-staticity depends on the holonomy and on the
distance between the constituents inside the caloron. 
For maximally nontrivial holonomy (which coincides with the average holonomy 
in the confinement phase) the constituent dyons have equal mass. 
For this simplified case we have determined a bifurcation value 
of the non-staticity, $\delta_t^{*}=0.27$, choosing the distance between the 
constituents such that the two lumps of action density (dyons) merge into one lump 
of action density. This single lump is what we call a ''recombined caloron''.
As already mentioned, the non-staticity uses the discreteness of a lattice configuration.
\begin{figure}
\includegraphics[width=.45\textwidth]{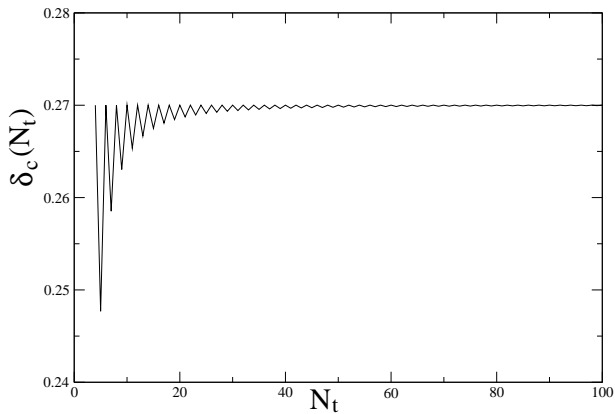}
\caption{The dependence of non-staticity $\delta_{t}^{*}$, defined at the bifurcation
point for analytic KvB caloron solutions, on the number of time slices $N_t$.
The zigzag form of the curve can be explained by the qualitatively different
arrangement of even and odd time points where the continuum KvB caloron solution
has to be calculated in order to evaluate (\ref{eq:nonst2}).}
\label{fig:ntlimit}
\end{figure}
The value $\delta_t^{*}$ is obtained by inserting the analytic form of the action 
density of a continuum KvB caloron solution~\cite{KvB}, calculated exactly at the point
of bifurcation, into the evaluation of (\ref{eq:nonst2}) using a grid with lattice 
spacing $a=b/N_t$. In order to see how well-defined at finite $N_t$ this ''bifurcation 
value'' $\delta_t^{*}$ can be, we have evaluated it for various (even and odd) 
$N_t \geq 4$. The function $\delta_t^{*}(N_t)$ that goes to $\delta_t^{*}=0.27$ for 
$N_t\rightarrow \infty$ is presented in Fig. \ref{fig:ntlimit}.

If the non-staticity is lower than the bifurcation value $\delta_t^{*}$  
two dyons can be distinguished by the two maxima of the action density. 
If the non-staticity is bigger than $\delta_t^{*}$ two dyons appear recombined 
into a caloron with only one action density maximum.

\section{Recombination of dyons into calorons with lowering temperature}
\label{sec:recombination}
The main results of this paper are obtained from ensembles of $SU(2)$ gauge field
configurations created by heat bath Monte Carlo at $\beta=2.2$ with
respect to the Wilson action $S_W$ on lattices $16^3 \times N_t$ with $N_t=4$,
$N_t=5$ and $N_t=6$.
Each ensemble consisted of about $8000$ independent configurations .
Cooling of these configurations was performed using the
fastest possible relaxation with respect to the Wilson action. For this
method each link $U_{x,\mu}$ is immediately replaced by the projection to
$SU(2)$ of the staples around it, $\tilde{U}_{x\mu}$.

The cooled configurations studied in this paper have been identified when
the cooling history finally has arrived at a quasi-stable plateau on the level of
a single instanton action $S_W \approx S_{inst}=2 \pi^2 \beta$.
Lateron, we describe also some results concerning
higher plateaux of action, e.g. $S_W \approx 3 S_{inst}$
which are passed earlier in the cooling process.
More precisely, on such plateaux we stopped cooling always at local minima of
the violation of the lattice equations of motion~\cite{IMMPSV}.
Here, the violation is defined as
\eq
 V = \sum_{x\mu} \left( \frac{1}{2} {\mathrm tr}~
 [ U_{x\mu} - \tilde{U}_{x\mu} ]^{\dagger}~
 [ U_{x\mu} - \tilde{U}_{x\mu} ] \right)^{1/2} \; .
 \label{eq:violation}
\en
In addition to that, the following conditions have been imposed for the
automatized selection of the classical solutions:
\begin{itemize}
\item the action fits into the window $~0.5 < S_W/S_{inst} < 1.25$,
\item the decrease of action has slowed down to $~|\Delta S_W|/S_{inst} < 0.05~$
and
\item the violation of the equations of motion should be sufficiently weak, $~V < 25$.
\end{itemize}
In most of the cases for which the first and second conditions hold we find a small minimum
violation $~V < 20$. In a few cases there were also local minima with much larger $V$-values,
which hardly could be interpreted as extended solutions of the lattice equations of
motion. Thus, the third condition was implied in order to select ``good'' solutions. We have
varied our stopping criteria~\footnote{The criteria have been tested for
$N_t=4$ and applied also to $N_t=5$ and $6$ (at the same $\beta$).}
and seen that the details of the ensemble of the 'frozen' objects slightly change.
Our main observations reported below do not depend on the details of the procedure.
The efficiency of the conditions was such that $80$ \% (in the case $N_t=4$),
$60$ \% ($N_t=5$) and  $55$ \% ($N_t=6$) of the equilibrium configurations
ended up in a cooled configuration at the one-instanton action plateau.
These cooled configurations include 
events with topological charge $Q = \pm 1$, {\it i.e.} real calorons, as well as 
configurations with topological charge $Q = 0$ which in our previous
paper \cite{IMMPSV} had been identified as static dyon-anti-dyon pairs ($D\bar D$). The 
latter constitute $18$ \% ($N_t=4$), $14$ \% ($N_t=5$) and $8$ \% ($N_t=6$), 
respectively, of all cooled configurations at the one-instanton level. They   
have been discarded from the considerations in this paper.  

We remind the reader that the recombination threshold $\delta_t^{*}=0.27$, strictly
speaking, reflects the recombination for maximally non-trivial holonomy only,
{\it i.e.} with an asymptotic value of the Polyakov line $L_{as}=0$. If one
performs cooling without special restrictions concerning the holonomy, there is
no guarantee that the asymptotic holonomy of the caloron configurations still
coincides with the average Polyakov line of equilibrium configurations in 
confinement. For the purpose of defining an {\it asymptotic} holonomy $L_{as}$ 
for each cooled configuration, we have determined the average of $L_{\vec x}$ 
over a $3D$ subvolume where the local $3D$ action density $s_{\vec x}$ is low, for 
definiteness $s_{\vec x} < 0.0001$.
\begin{figure}
\includegraphics[width=.40\textwidth,angle=-90]{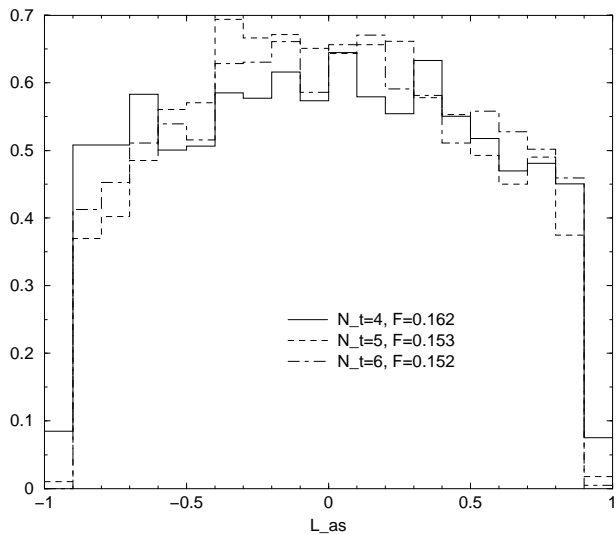}
\caption{The distribution of holonomy $L_{as}$ for the three samples
of cooled configurations corresponding to three temperatures below
deconfinement at $N_t=4$, $N_t=5$ and $N_t=6$.}
\label{fig:hiltop}
\end{figure}
In Fig. \ref{fig:hiltop} we present the distribution of cooled configurations over
$L_{as}$ as a histogram (with bin size 0.1) for the three cases $N_t=4$, $N_t=5$ 
and $N_t=6$. In the legend we show the respective volume fraction ($F \approx 0.15$)
of the three-volume 
over which the ''asymptotic'' value $L_{as}$ is defined as an average,
{\it i.e.} far from the lumps of action and topological charge. 

As explained above, the non-staticity $\delta_t$ is a measure which describes 
the distance from a perfectly (Euclidean) time independent configuration. In other 
words, the distributions of non-staticity of caloron events obtained by cooling 
of equilibrium lattice configurations can be considered as a substitute for the
distribution in dyon distances $d$. This quantity can be directly measured 
for cooled lattice gauge field configurations.
\begin{figure*}
\centering
\includegraphics[width=.40\textwidth,angle=-90]{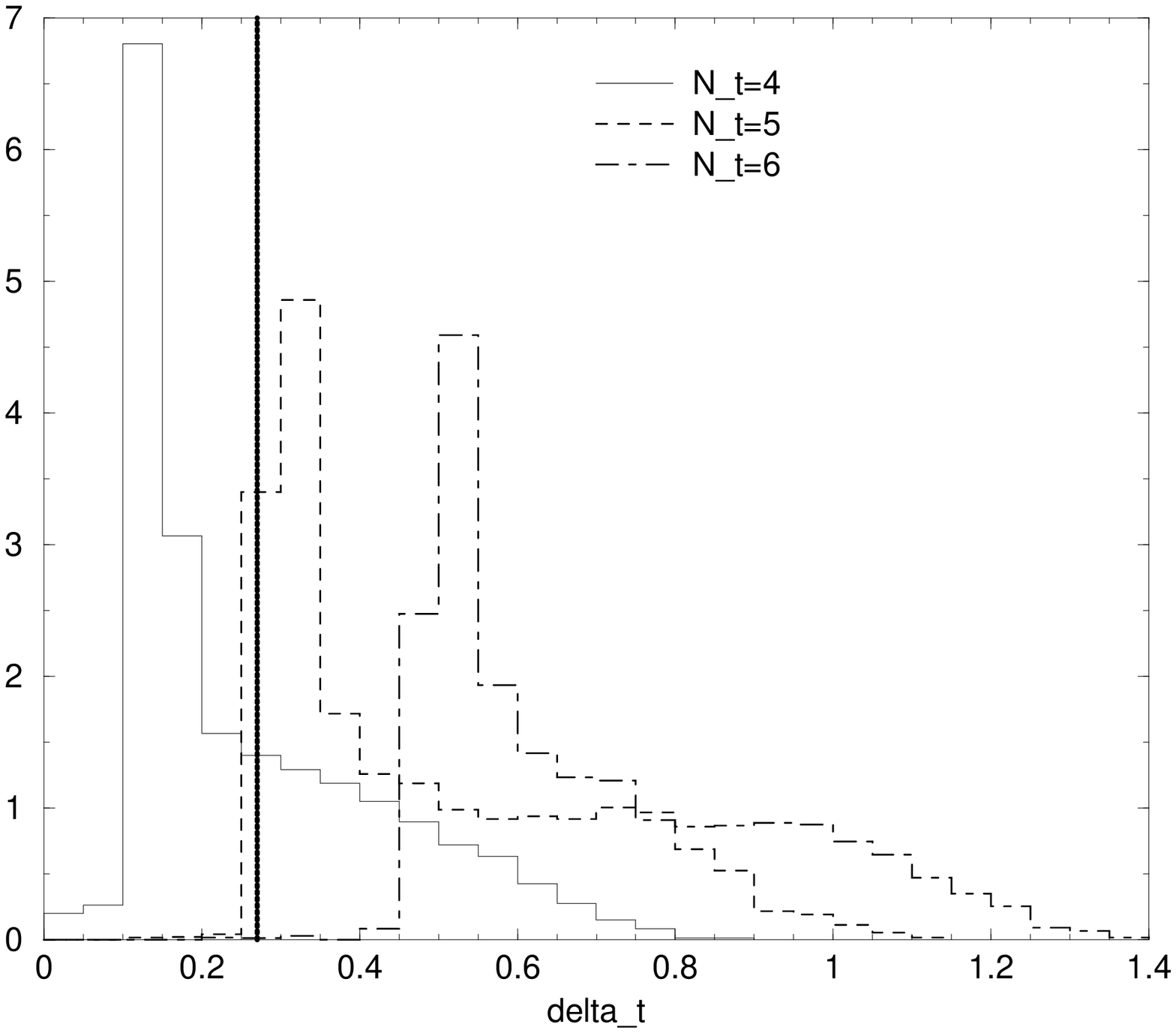}\qquad%
\includegraphics[width=.40\textwidth,angle=-90]{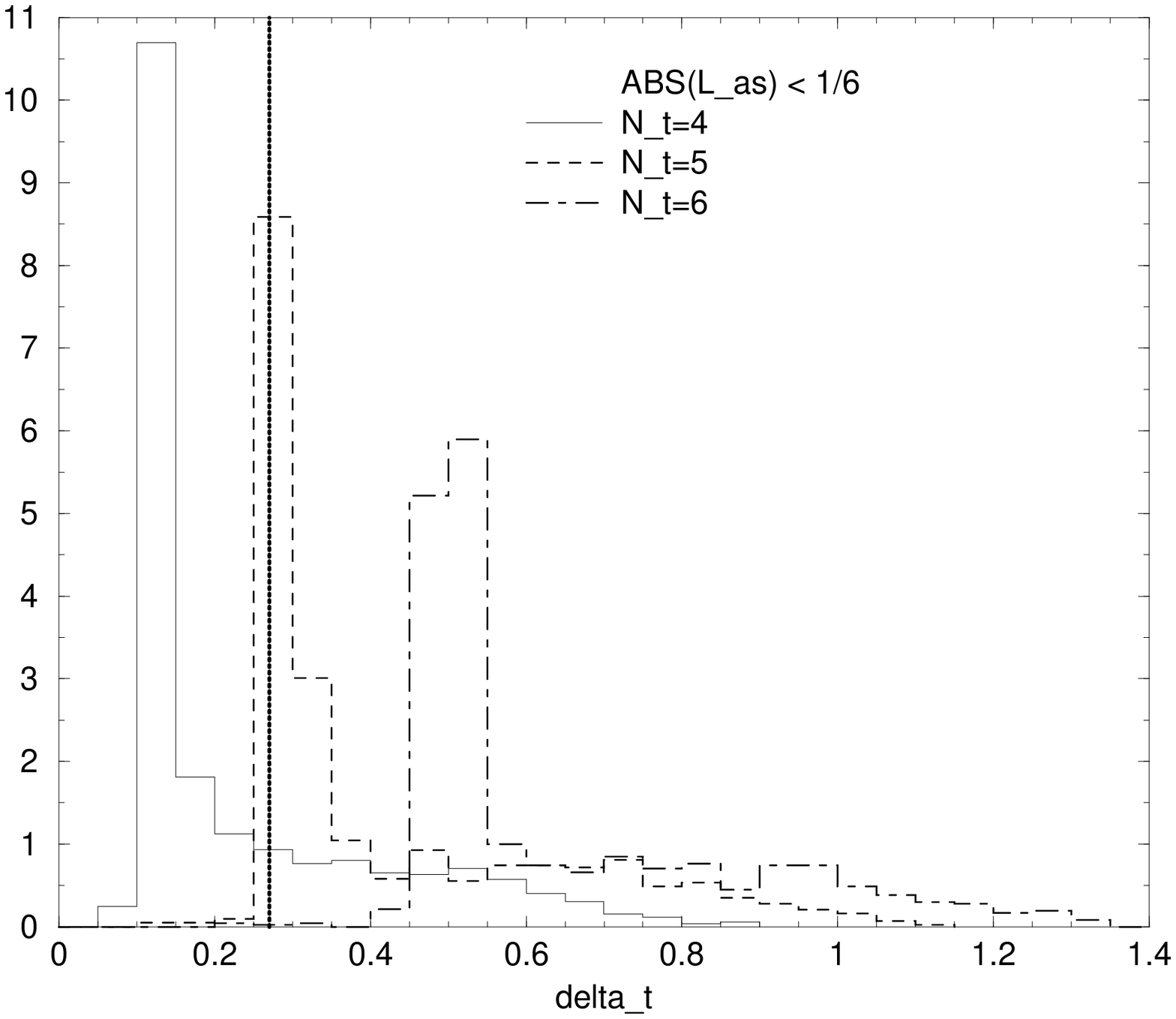}

(a) ~~~~~~~~~~~~~~~~~~~~~~~~~~~~~~~~~~~~~~~~~~~~~~~~~~~~~~~~~~~~~~~~~~~~~~~~~~ (b)
\caption{The distribution of non-staticity $\delta_t$ after cooling
as histograms (with bin width 0.05), for three temperatures below
deconfinement at $N_t=4$, $N_t=5$ and $N_t=6$ :
(a) without any cut,
(b) when a cut $|L_{as}|<1/6$ is applied.
The thick vertical line marks the non-staticity $\delta_{t}^{*}$
where the caloron recombines.}
\label{fig:t456}
\end{figure*}
We show in Fig. \ref{fig:t456}
the $\delta_t$ distributions for all our cooling products obtained at 
$\beta = 2.2$ on $16^3 \times N_t$ lattices.

In an attempt to make a fair comparison with calorons with non-trivial holonomy
and to correct for the possible evolution of the asymptotic holonomy away from 
$L_{as}=0$ during the cooling process, we defined a subsample by the requirement 
$|L_{as}| < 1/6$. One can see that the cut with respect to the asymptotic holonomy 
selects cooled configurations from the flat central
part of the histogram shown in Fig. \ref{fig:hiltop}. On the other hand, we notice
that a considerable fraction of cooled configurations has developed an asymptotic 
holonomy $|L_{as}| > 1/6$.

In Fig. \ref{fig:t456} (a) we show the probability distribution over $\delta_t$ for 
cooled configurations with an action at the one-instanton plateau without the
cut according to the asymptotic holonomy $|L_{as}|$. One can see that a relatively 
high fraction of configurations, obtained from the Monte Carlo equilibrium
with $N_t=4$, has $\delta_t < \delta_t^{*}=0.27$. This means that they would be 
identifiable as consisting of two constituents by looking for the $3D$ action density 
on the lattice. For $N_t=5$ it is only a minority of cooled configurations which 
falls below the threshold $\delta_t=0.27$. No static (according to the non-staticity 
criterum) configurations have been found among cooling products at $N_t=6$.

We have repeated the same analysis after applying the cut with respect to the 
asymptotic holonomy, $|L_{as}|<1/6$. Then we get modified histograms in $\delta_t$ 
for the three temperatures. 
This is shown in Fig. \ref{fig:t456} (b). The histograms got more pronounced peaks 
in $\delta_t$ which are positioned around 0.125, exactly around $\delta_t^{*}=0.27$, 
and around 0.5 for $N_t=4$, $N_t=5$ and $N_t=6$, respectively.  

There are other criteria which could be used to characterize a more or less 
static configuration, for example, the presence of static Abelian monopoles 
emerging in the maximal Abelian projection. This criterium is not identical 
with the separation set by $\delta_t^{*}$. 
Thus, another subsample can be defined by the property that a pair of static
Abelian monopoles has been found after fixing the cooled configuration 
to the maximal Abelian gauge and doing the Abelian projection.
This subsample can be analyzed with respect to the 
three-dimensional distance $R$ between the Abelian monopoles.

\begin{figure*}
\centering
\includegraphics[width=.40\textwidth,angle=-90]{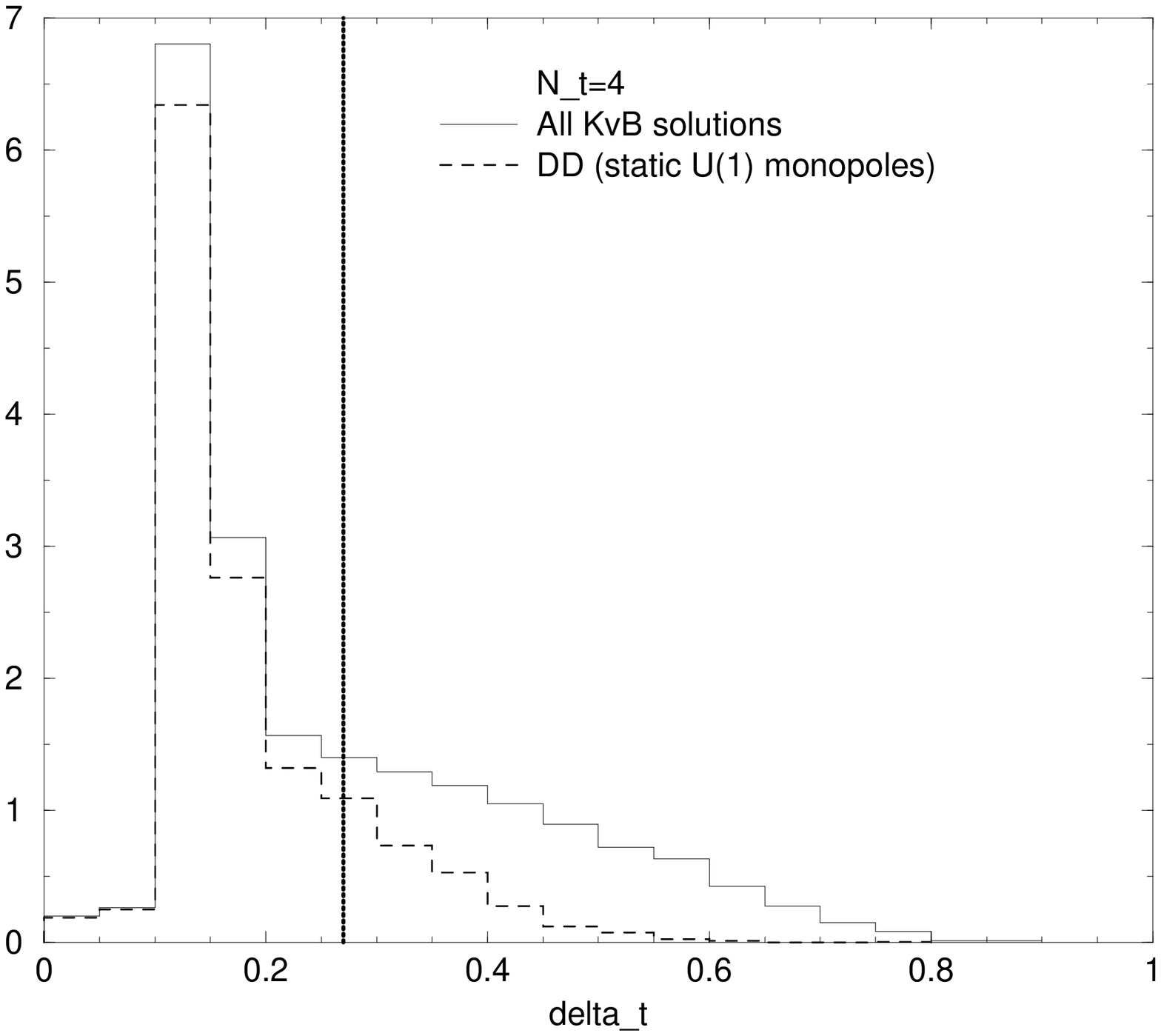}\qquad%
\includegraphics[width=.40\textwidth,angle=-90]{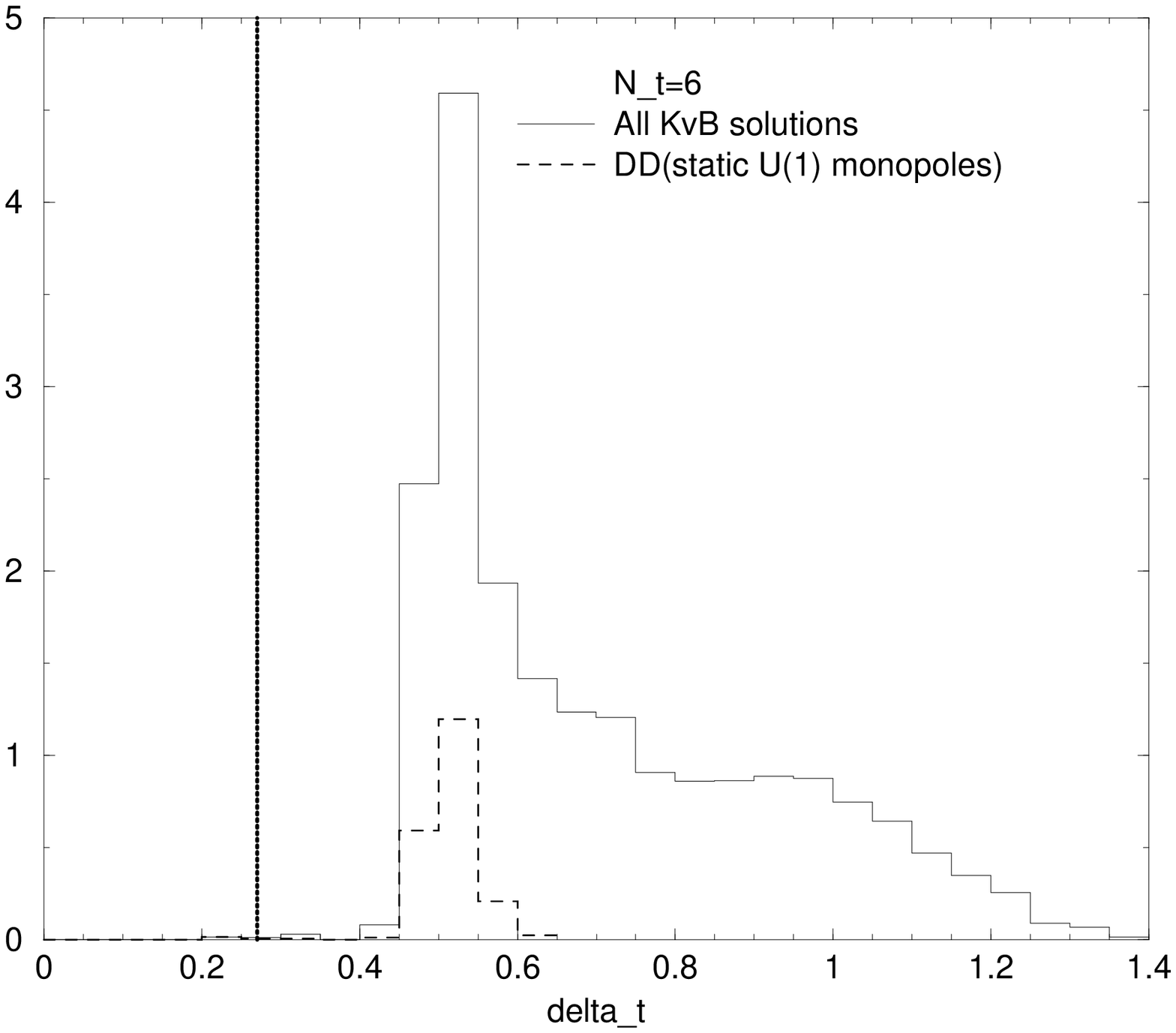}

(a) ~~~~~~~~~~~~~~~~~~~~~~~~~~~~~~~~~~~~~~~~~~~~~~~~~~~~~~~~~~~~~~~~~~~~~~~~~~ (b)
\caption{The distribution of non-staticity $\delta_t$ after cooling
compared with the subsample of configurations which have a static
dyon-dyon (DD) structure exhibited by monopoles after Abelian projection:
(a) for the higher temperature with $N_t=4$, (b) for the lower
temperature with $N_t=6$.
The thick vertical line marks the non-staticity $\delta_{t}^{*}$
where the caloron recombines.}
\label{fig:t46}
\end{figure*}
In Fig. \ref{fig:t46} (a) we show the histogram over $\delta_t$ of all cooled 
configurations obtained from the Monte Carlo ensemble at $N_t=4$ together 
with the histogram of those which explicitely exhibit the dyon-dyon structure
in terms of Abelian monopoles. One can see that practically all cooled 
configurations below $\delta_t^{*}$ possess this structure, but above $\delta_t^{*}$ 
the fraction rapidly goes to zero.
We show the same for $N_t=6$, {\it i.e.} at lower temperature, in Fig. \ref{fig:t46} (b).
In this case, at the peak value around $\delta_t=0.5$ only 20 \% of the solutions 
still are characterized by a static dyon-dyon pair, whereas at higher non-staticity
this is never the case.
In these two distributions no cut with respect to the asymptotic holonomy has been
applied.

Up to now we have two tentative definitions of the position of the constituents:
one is the position of the 
two static Abelian monopoles (in MAG, as long as they they are static) and the other is 
defined by the maxima of the $3D$ action density.~\footnote{For the case of analytical
caloron solutions, the relation between the constituent locations, the locations of
the maxima of the action density and the locations where $|P|=1$ has been compared
in Ref.~\cite{JHEP}.} 
In the case of well-separable maxima of the latter these maxima fall close to
the positions of the MAG monopoles. In the other limiting cases of recombined 
maxima or 
imbalanced maxima (this corresponds to an asymptotic holonomy far from zero)  
of the $3D$ action density at least the {\it absolute maximum of action density} 
is still easy to find. It is 
either {\it the} single maximum or it takes the role of {\it one} of the maxima.
\begin{figure*}
\centering
\includegraphics[width=.40\textwidth,angle=-90]{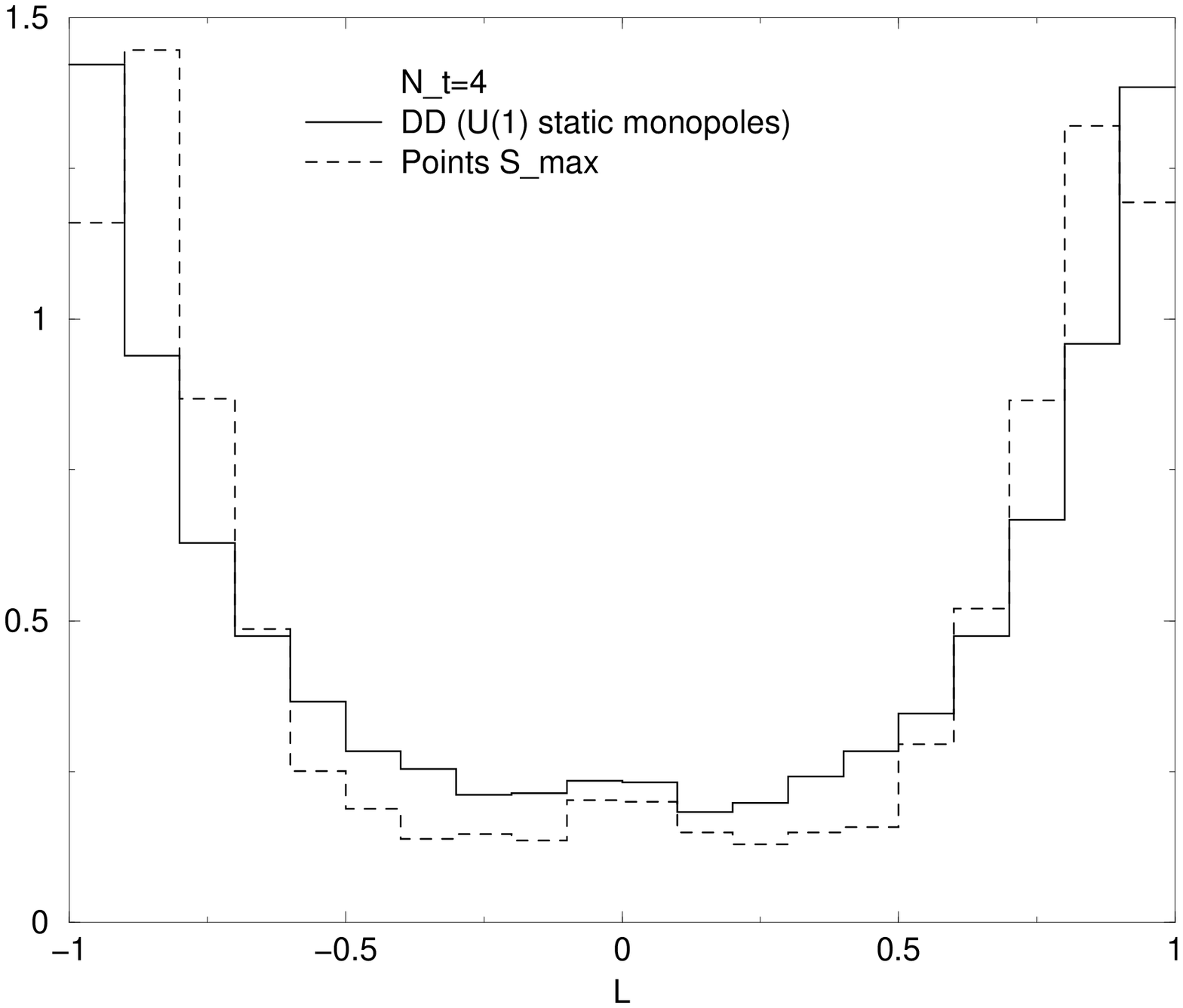}\qquad%
\includegraphics[width=.40\textwidth,angle=-90]{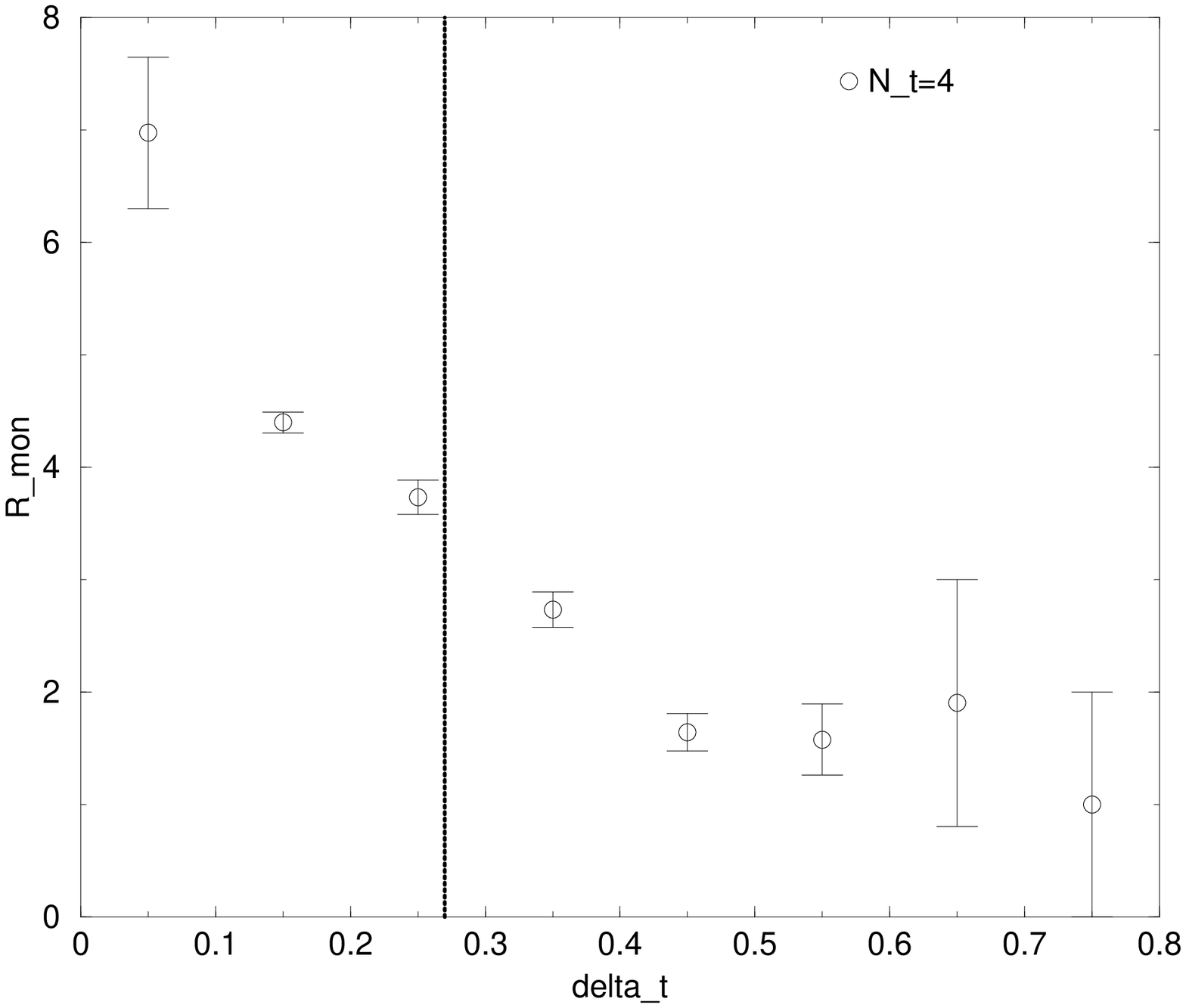}

(a) ~~~~~~~~~~~~~~~~~~~~~~~~~~~~~~~~~~~~~~~~~~~~~~~~~~~~~~~~~~~~~~~~~~~~~~~~~~ (b)
\caption{The dyon-dyon structure after Abelian projection :
(a) distribution of the local Polyakov loop at the position of
the static Abelian monopoles compared to the distribution
at the maxima of action density,
(b) average distance between the static Abelian monopoles vs.
non-staticity $\delta_t$ assigned to the cooled configuration.
The thick vertical line marks the non-staticity $\delta_{t}^{*}$
where the caloron recombines.}
\label{fig:histl1}
\end{figure*}
For the case of $N_t=4$, {\it i.e.} the temperature just below deconfinement,
we show in Fig. \ref{fig:histl1} (a) the histogram with respect to the local 
Polyakov loop at the two sorts of $3D$ constituent points, the loci of static 
monopoles or the maximum of action density. In both definitions the histogram 
peaks near to $L_{\vec x} = \pm 1$. The peak is however more pronounced for the 
monopole locations, less pronounced for the maxima of the $3D$ action density.

The relation between the non-staticity $\delta_t$ and the distance $R$ 
between the static dyonic constituents emerging (in MAG as Abelian monopole) 
given in lattice units, is presented in Fig. \ref{fig:histl1} (b) for the higher
temperature, near the deconfinement temperature ($N_t=4$). For this temperature
such dyon-dyon events are clearly distinguishable among the cooled configurations 
as long as the non-staticity $\delta_t < 0.6$. For extremely low non-staticity 
$\delta_t$ (left from the peaks in Fig. \ref{fig:t456} (a) and (b)) we find an 
average distance $R \approx 7$, whereas near $\delta_t^{*}$ the average distance 
is $R \approx 4$. For higher $\delta_t$, the part of solutions which still possesses 
a clear dyon-dyon structure in terms of Abelian monopoles, has them localized at 
distances $R$ between one and two lattice spacings.

In order to represent how the Polyakov line behaves inside a lump of action,
we have chosen the absolute maximum of the $3D$ action density 
(denoted as central point ${\vec x}_0$) 
and have explored the Polyakov loop in its neighborhood. 
For this purpose, we have defined a locally summed-up Polyakov line 
$L_{tot}$ (summed over the central point ${\vec x}_0$ and its six nearest
neighbors ${\vec x}_i$ ($i=1,..,6$): 
\eq
 L_{tot} = \sum_{i=0}^{6} L_{{\vec x}_i}
 \label{eq:summed_L}
\en
and a kind of Polyakov-line dipole moment over the same set of $3D$ lattice
points with respect to the central point:
\eq
 {\vec M}_{tot} = \sum_{i=1}^{6} L_{{\vec x}_i} ({\vec x}_i - {\vec x}_0) \; . 
 \label{eq:dipole_M}
\en
The absolute value $|L_{tot}|$ of the first quantity tests the amount of 
local coherence of the Polyakov line. The absolute value $|{\vec M}_{tot}|$
of the second quantity tests the amount of presence of opposite-sign Polyakov 
lines representing eventually two different constituents inside the same lump 
of action.
\begin{figure*}
\centering
\includegraphics[width=.40\textwidth,angle=-90]{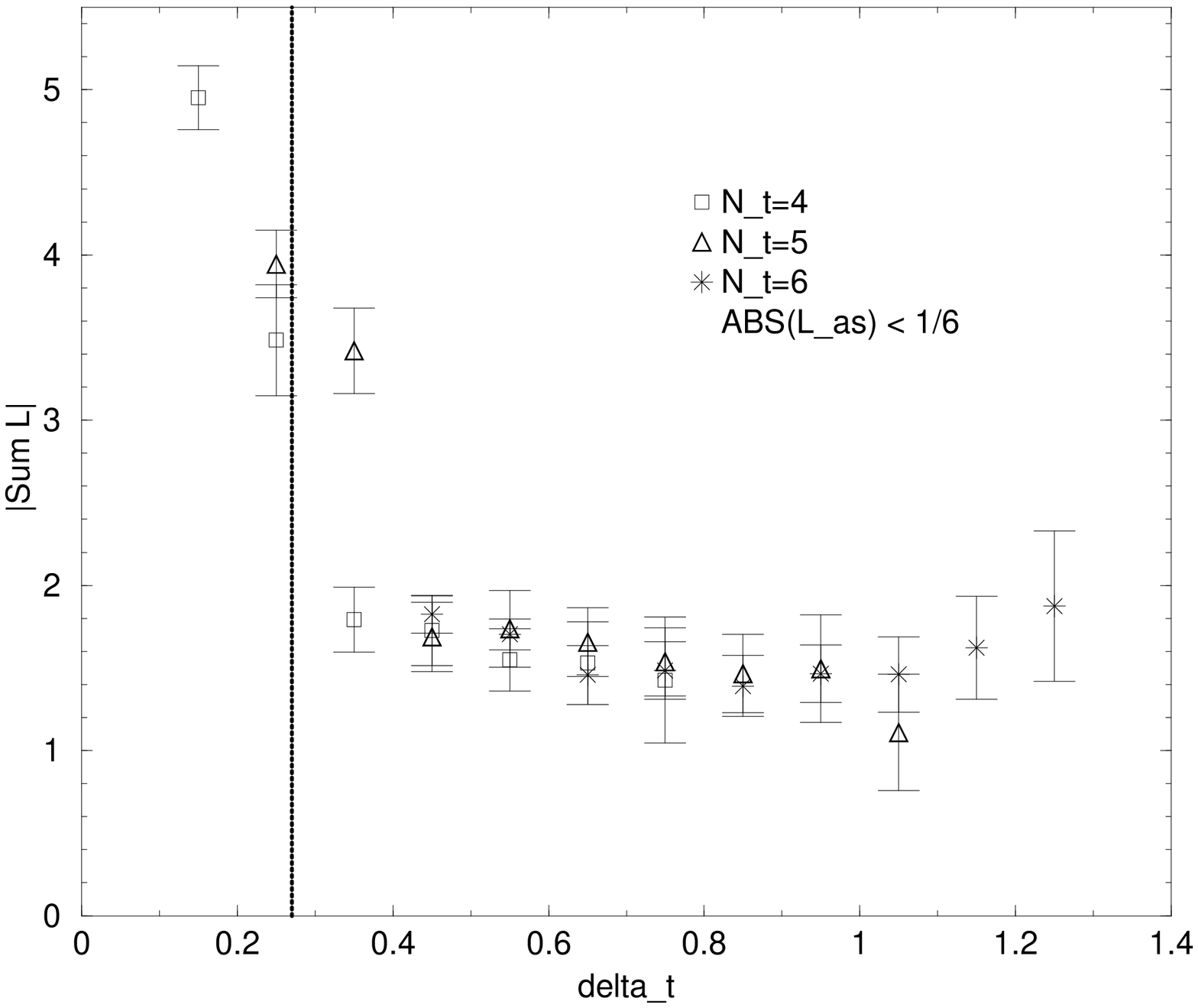}\qquad%
\includegraphics[width=.40\textwidth,angle=-90]{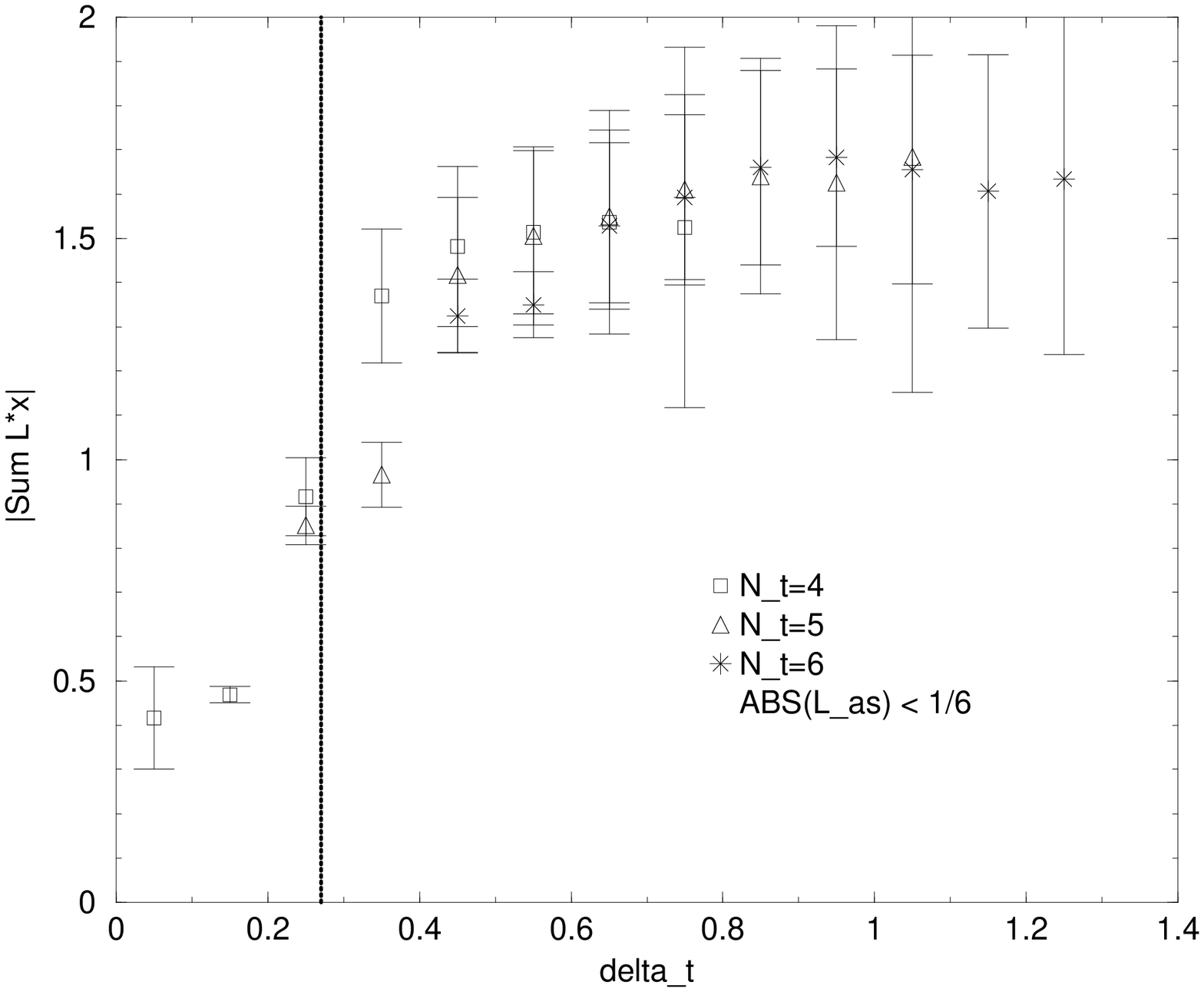}

(a) ~~~~~~~~~~~~~~~~~~~~~~~~~~~~~~~~~~~~~~~~~~~~~~~~~~~~~~~~~~~~~~~~~~~~~~~~~~ (b)
\caption{The Polyakov line in the neighborhood of maxima of the
action density :
(a) modulus of the average summed-up Polyakov line 
$|L_{tot}|$ and (b) modulus of the corresponding ''dipole moment'' 
$|{\vec M}_{tot}|$ vs. non-staticity $\delta_t$,
for the subsample with asymptotic holonomy near zero. For
details of the definition see the text. 
The thick vertical line marks the non-staticity $\delta_{t}^{*}$
where the caloron recombines.}
\label{fig:qt1w}
\end{figure*}
Fig. \ref{fig:qt1w} (a) shows how the $|L_{tot}|$ ({\it i.e.} the locally summed-up 
Polyakov line) changes with $\delta_t$ in different bins of width 0.1. 
For the temperature nearest to the transition, at $N_t=4$, we see that $|L_{tot}|$
falls from $\approx 4.0$ to $\approx 1.0$ at $\delta_t \ge 0.5$. We interprete 
this such that in the region, where constituents can be well separated according 
to the action density (at small $\delta_t$), they are characterized by a relatively 
smooth change of the Polyakov line inside. In the region of large $\delta_t$ where 
they are not separable according to the action density, the Polyakov line changes 
rapidly in the neighbourhood of the absolute maximum of action density. 
For the lower temperatures, $N_t=5$ and $N_t=6$, the relationship between these 
properties of an action cluster and $\delta_t$ is the same.
The difference is that separable lumps of action 
(those with low $\delta_t$) become very rare.
Fig. \ref{fig:qt1w} (b) shows how the ''dipole moment'' $|{\vec M}_{tot}|$ of the 
Polyakov line around a maximum of action density rises with increasing 
non-staticity $\delta_t$. In the region where one can separate the constituents 
according to the action density, the dipole moment is small, emphasizing again 
the homogeneity of the Polyakov line around the central point. In the region beyond 
$\delta_t^{*}$ the dipole moment gradually stabilizes around a value of 1.5.

\begin{figure}
\includegraphics[width=.40\textwidth,angle=-90]{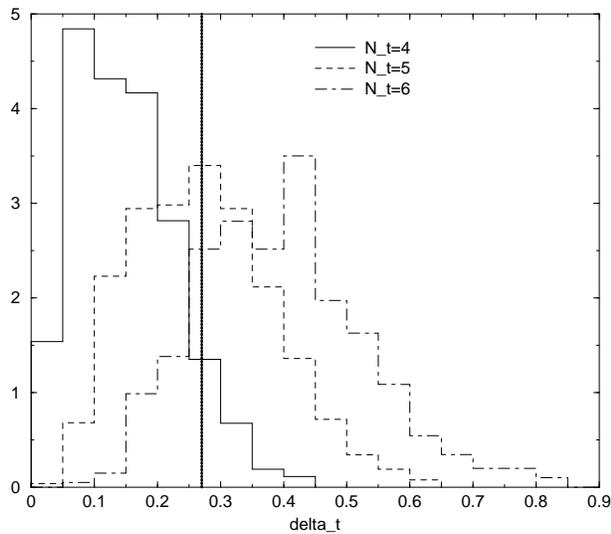}
\caption{The same as in Fig. 3 (a) for $S=3 S_{inst}$ plateaux.
No cut in the asymptotic holonomy has been applied.
The thick vertical line marks the non-staticity $\delta_{t}^{*}$
where the caloron recombines.}
\label{fig:p3t456}
\end{figure}
In the same way as described so far, we have analyzed configurations obtained by 
cooling at higher action plateaux. As an example we show in Fig. \ref{fig:p3t456} 
the histogram of non-staticity $\delta_t$ for the same three temperatures 
represented by $N_t=4$, $N_t=5$ and $N_t=6$. Although the precise border between
static and non-static has not the clear meaning as for the one-caloron case, the 
trend is the same: at lower temperature the lumps of action tend to be more 
localized also in Euclidean time (''instanton-like''). Compared with the 
one-caloron case, the histograms are more smeared. 

So far we have investigated the outcome from cooling for varying temporal lattice
extent and fixed spatial volume. What happens when changing $~\beta~$ at fixed
lattice size and fixed lattice asymmetry ?
\begin{figure*}
\centering
\includegraphics[width=.40\textwidth,angle=-90]{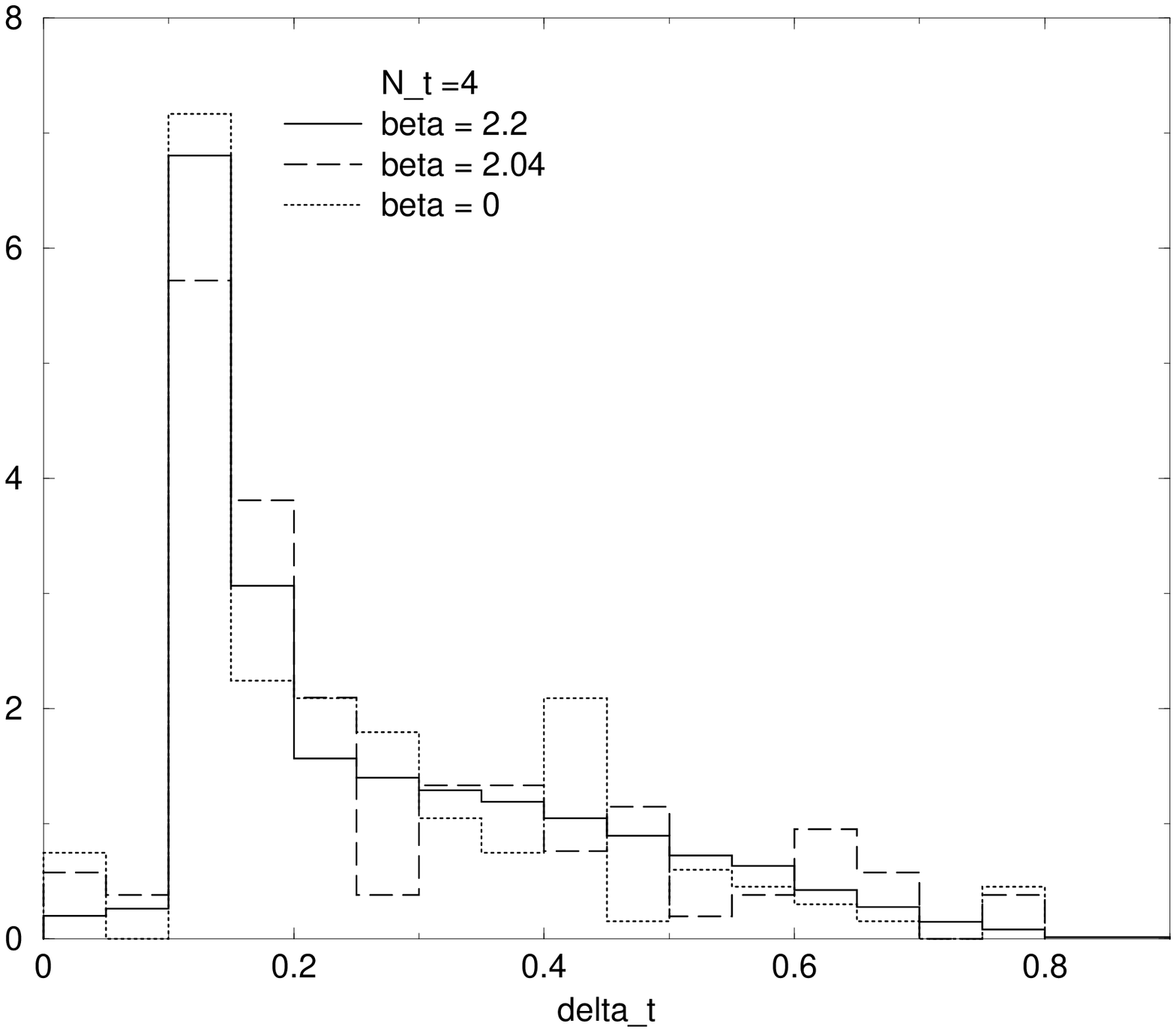}\qquad%
\includegraphics[width=.40\textwidth,angle=-90]{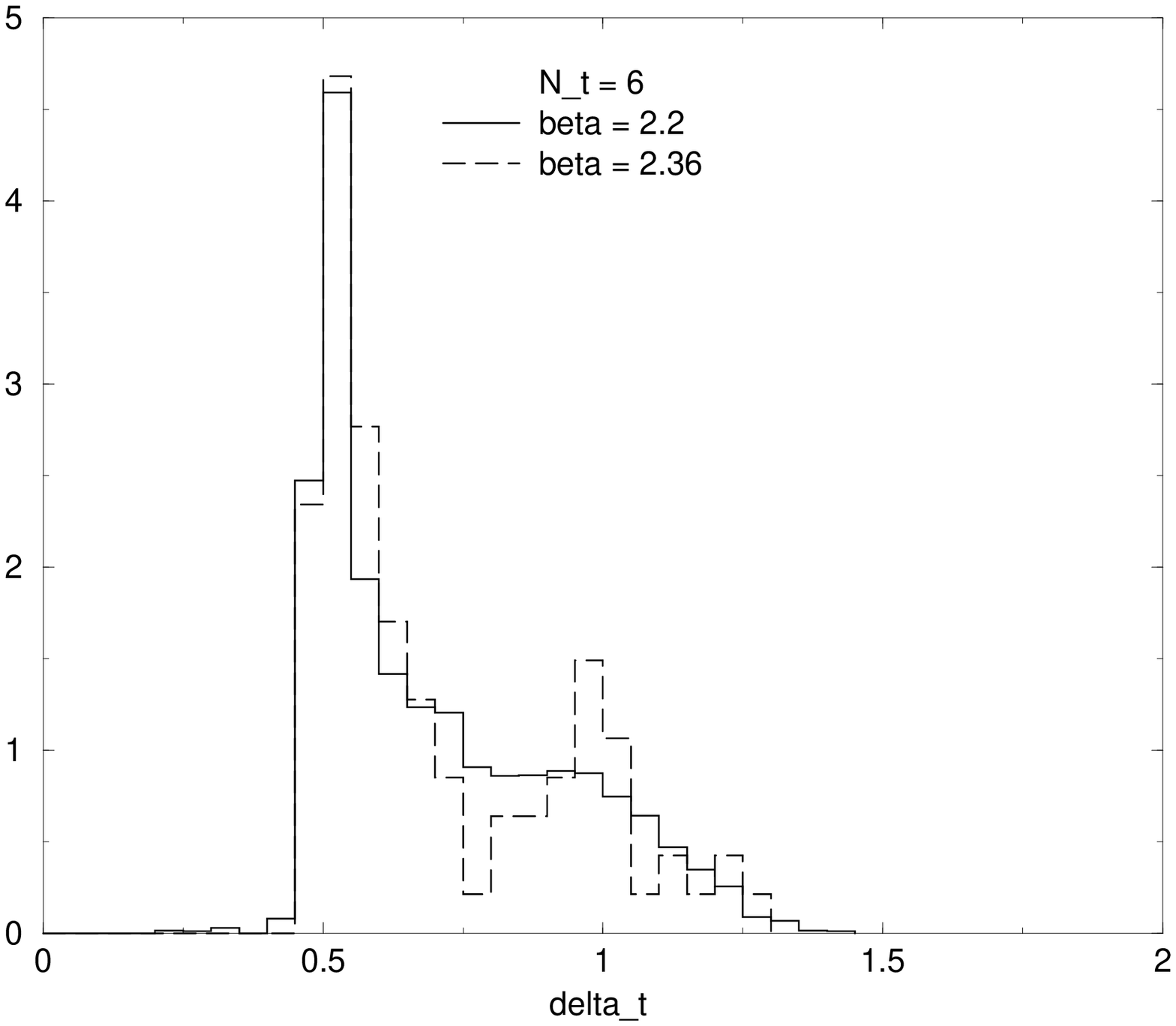}

(a) ~~~~~~~~~~~~~~~~~~~~~~~~~~~~~~~~~~~~~~~~~~~~~~~~~~~~~~~~~~~~~~~~~~~~~~~~~~ (b)
\caption{The distribution of non-staticity $\delta_t$ after cooling for 
fixed $N_t$ but varying $\beta$-values. No cut in the asymptotic holonomy 
is made.  The spatial lattice size is always $16^3$.
(a) $N_t=4: ~\beta=0.0, 2.04, 2.20$, (b) $N_t=6:~\beta=2.20, 2.36$. }
\label{fig:scaling}
\vspace{0.2cm}
\end{figure*}
In Figs. \ref{fig:scaling} we again show
non-staticity $~\delta_t~$ histograms without applying cuts for $~|L_{as}|$.
We compare (a) for $~N_t=4~$ the cases $~\beta =0~$, $~\beta =2.04~$ 
and $~\beta=2.20~$ with each other, and
(b) for $~N_t=6~$ the case $~\beta = 2.20~$ with $~\beta = 2.36~$.
\footnote{Note that according to asymptotic scaling 
$\beta=2.20, N_t=4$ and $\beta=2.36, N_t=6$ 
would correspond approximately to the same physical temperature}
The statistics for the new $\beta$-values ($0.0, 2.04$ and $2.36$) is of 
order $O(200)$ field configurations each.
Contrary to what one might have expected, we see that among the cases (a) 
and the cases (b), respectively, there are no qualitative differences.
This means that the ratio of probabilities for the occurence of dissociated 
calorons (or separate static dyon pairs) {\it vs.} non-dissociated (single lump) 
calorons is {\bf not} a function of the temperature 
describing the original equilibrium ensemble.
Instead, it is mainly determined by the geometry (the aspect ratio) of the lattice. 
The larger the temporal lattice extent is in comparison with the spatial extent 
the lower is the probability to find separated lumps of more or less static 
objects after cooling. In other words, once there is only one topological 
$Q\ne0$ object left, everything is classical. As a rule, the object is smooth, 
such that only the size of the box matters or other "infrared forces" 
({\it e.g.} those forbidding single instantons) play a role.
This observation clearly shows the limitation of the cooling method applied to
the lowest action plateaux.  

\section{Instantons or calorons on a symmetric 4-torus}
\label{sec:torus}
\begin{figure*}
\centering
\includegraphics[width=.40\textwidth]{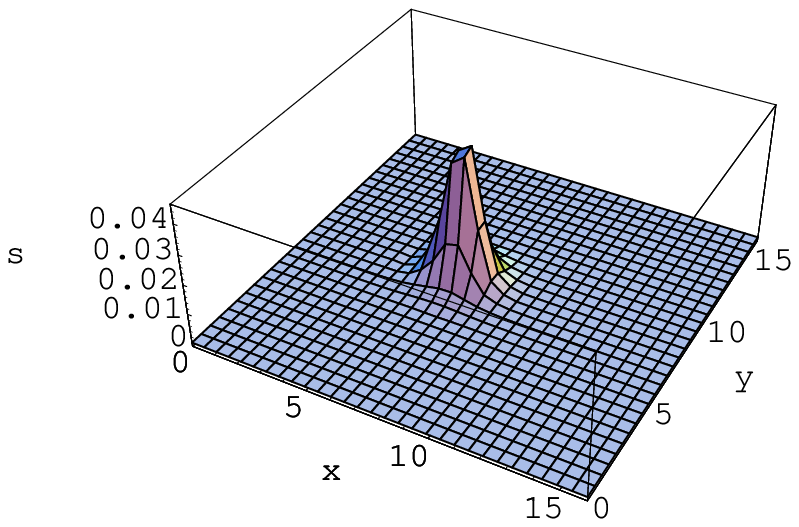}\qquad%
\includegraphics[width=.40\textwidth]{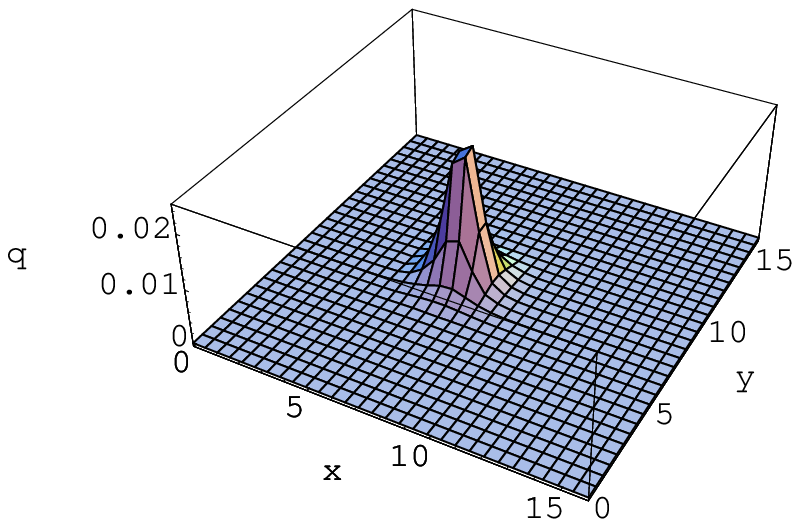}
\vspace{1.0cm}

\includegraphics[width=.40\textwidth]{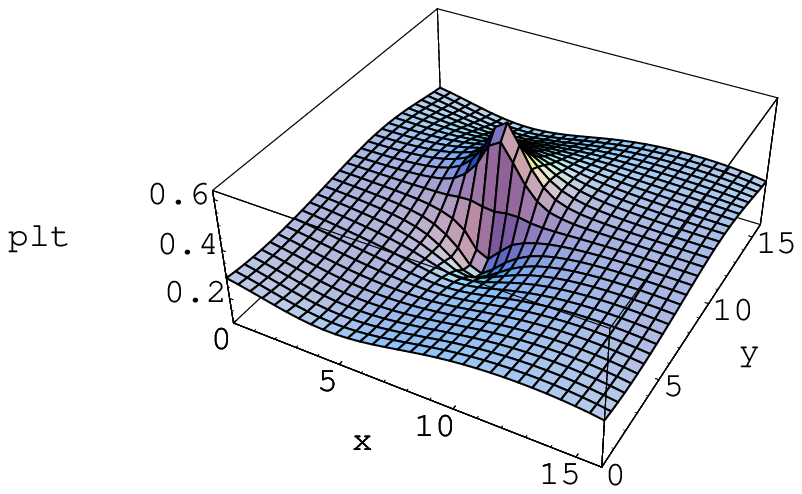}\qquad%
\includegraphics[width=.40\textwidth]{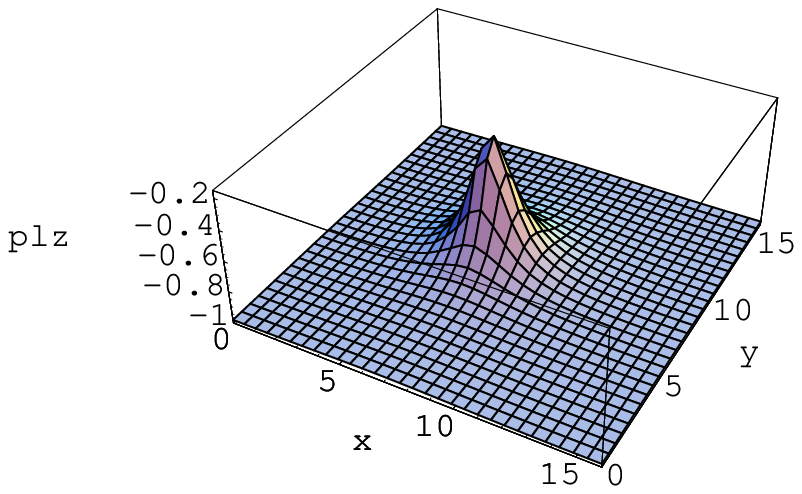}
    
\vspace{1.0cm}

\includegraphics[width=.40\textwidth]{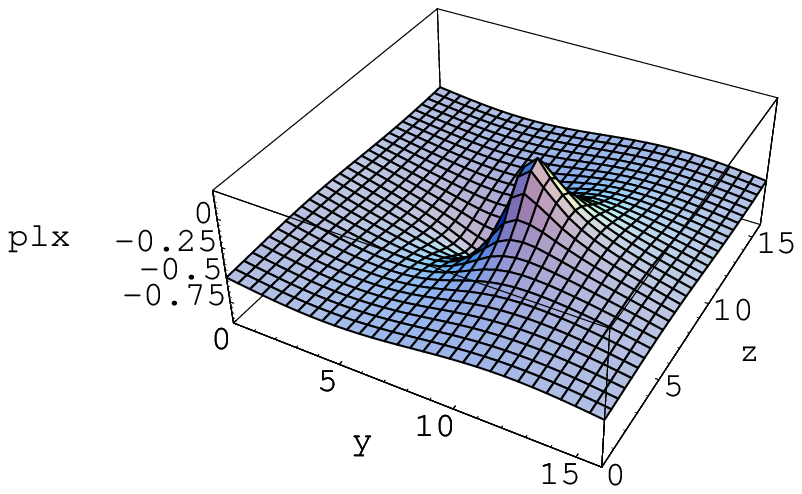}\qquad%
\includegraphics[width=.40\textwidth]{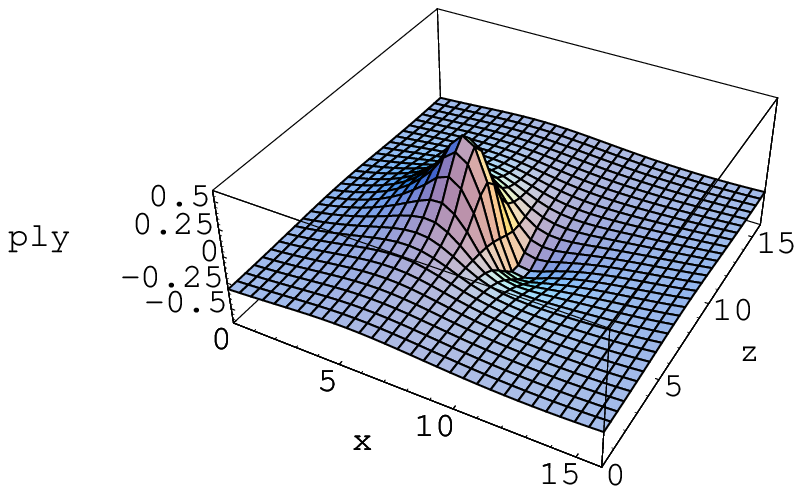}

\caption{Profiles of the action density ($s$), the topological charge density ($q$),
and  of the Polyakov lines ($plt,plx,ply,plz$) calculated along all straight
line pathes parallel to the four axes for a $16^4$ lattice caloron
found by cooling a Monte Carlo generated equilibrium gauge field down to
the one-instanton action plateau. The center of the caloron
(at the maximum of its action density) was found at the site
$(x,y,z,t)=(7,8,8,14)$. The planes shown in the figures cross just
this point.}
\label{fig:zerotemp}
\end{figure*}

With low statistics ($100$ configurations ) we have
also cooled equilibrium configurations generated with $\beta=2.2$
on symmetric lattices ($16^4$ representing ''zero'' temperature). 
In this case we have found for the classical configurations at the plateau 
$S_W \approx S_{inst}$ a broad distribution of non-staticity with a maximum
around $\delta_t \approx 2$ and with a tail extending beyond 3
\begin{figure}
\vspace{0.2cm}
\includegraphics[width=.45\textwidth]{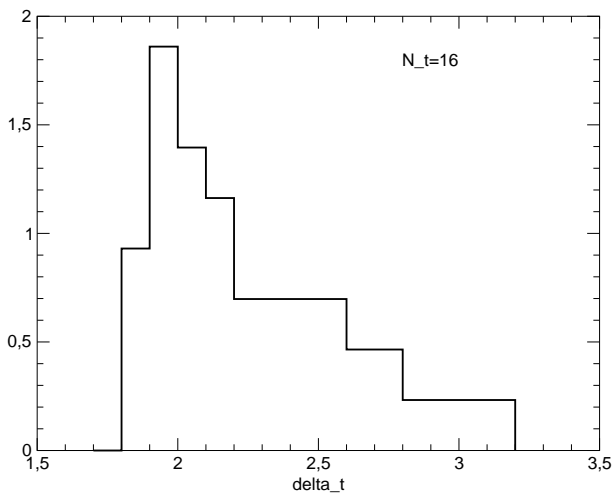}
\caption{The distribution of non-staticity $\delta_t$ after cooling for 
$\beta =2.2$ on the symmetric lattice $16^4$, obtained from $42$ 
configurations with unit topological charge.}
\label{nonstatzerotemp}
\end{figure}
(see Fig. \ref{nonstatzerotemp}). These are obviously
configurations with an action (topological charge) density well-localized 
in all four Euclidean directions.
There is a non-trivial behaviour of the Polyakov line inside these non-dissociated,
instanton-like objects resembling the Polyakov line associated with the time 
direction in the finite temperature case for non-dissociated calorons.
We have mapped the cooled lattice configuration 
with the help of all 4 possible definitions of the Polyakov line, 
which are now, on a symmetric lattice, physically equivalent to each other. 
Fig. \ref{fig:zerotemp} shows the 
profiles of action density, topological charge, and of the Polyakov lines (for 
four possible definitions) as they are seen in appropriate planes intersecting 
the lump through the centrum. The latter is defined as the maximum of the $4D$ 
action density. For all types of Polyakov lines the characteristic double structure
is seen exactly when the ''asymptotic'' value of the respective Polyakov line is 
{\it not close} to $\pm 1$. 

It is interesting to compare the pattern of the Polyakov line of these ''caloron 
candidates'' with the analytic KvB caloron formally constructed on a $16^4$ lattice 
corresponding to maximally non-trivial holonomy with respect 
to what has been chosen as ''time'' direction. 
For this construction, the two constituents have been placed along the 
$z$-direction, separated by 8 lattice spacings. 
The lattice caloron is obtained calculating link by link from the continuum gauge 
field $A_{\mu}$. Of course, such a constructed lattice caloron has irregularities 
at the boundary if the lattice action is evaluated under the assumption of periodicity. 
After some cooling the configuration turns into an (approximate) solution of the
lattice equations of motion {\it on the torus}. 
\begin{figure*}
\centering
\includegraphics[width=.40\textwidth]{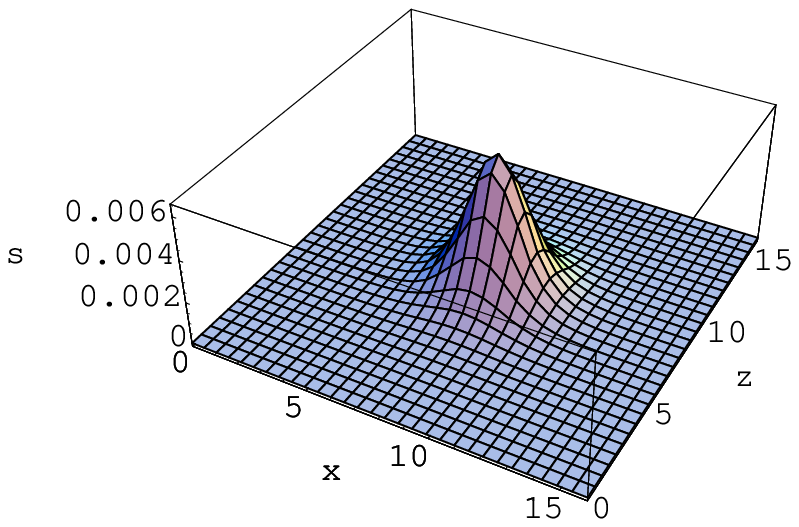}\qquad%
\includegraphics[width=.40\textwidth]{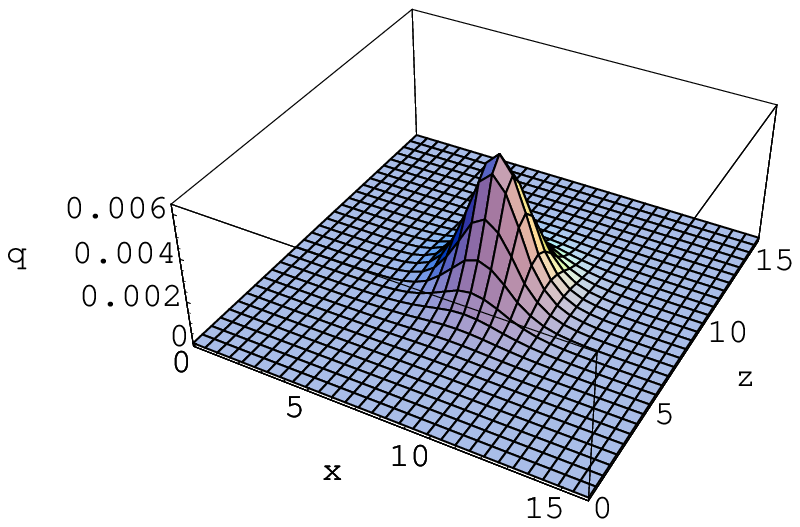}
    
\vspace{1.0cm}

\includegraphics[width=.40\textwidth]{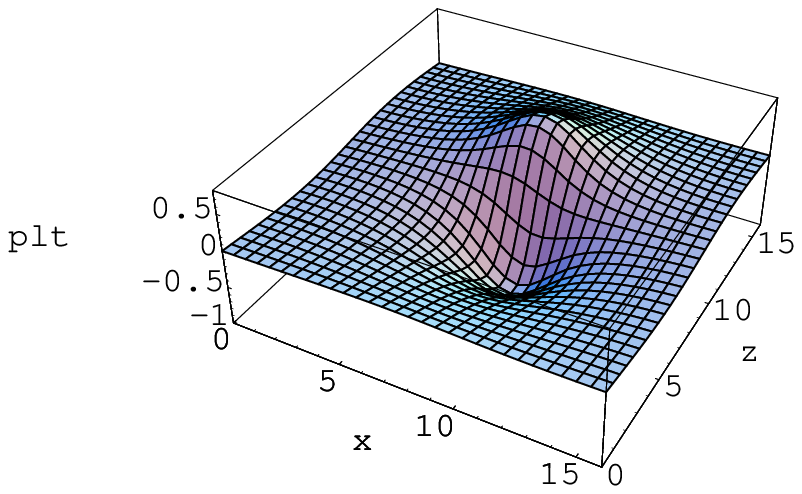}\qquad%
\includegraphics[width=.40\textwidth]{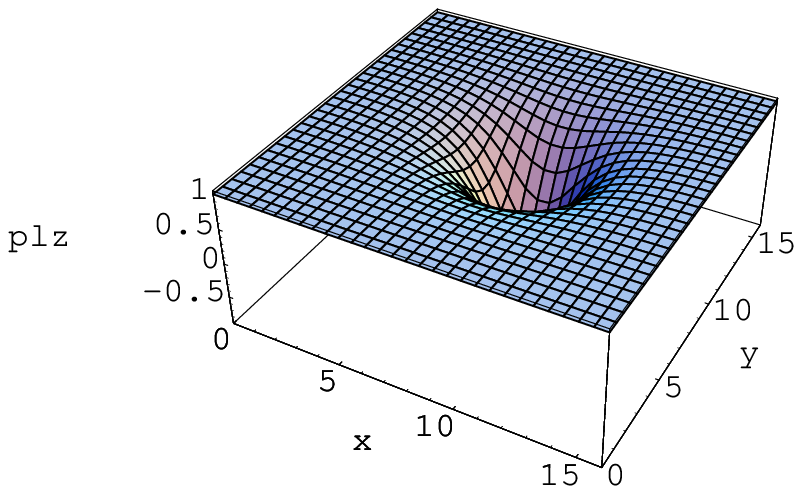}
    
\vspace{1.0cm}

\includegraphics[width=.40\textwidth]{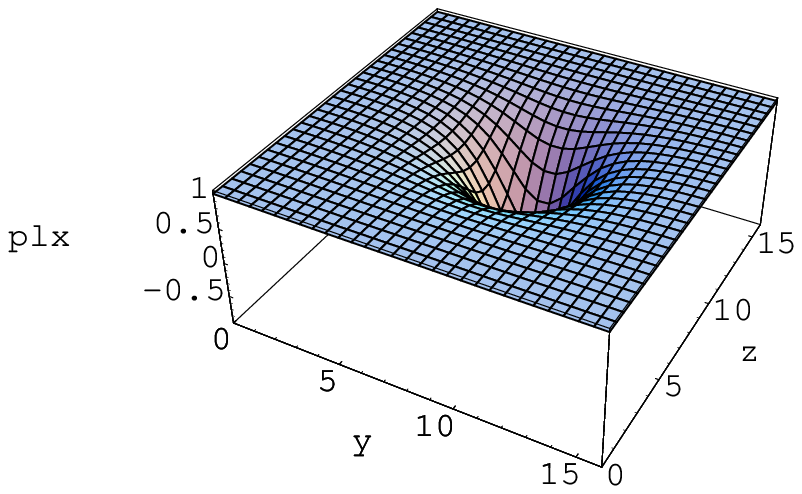}\qquad%
\includegraphics[width=.40\textwidth]{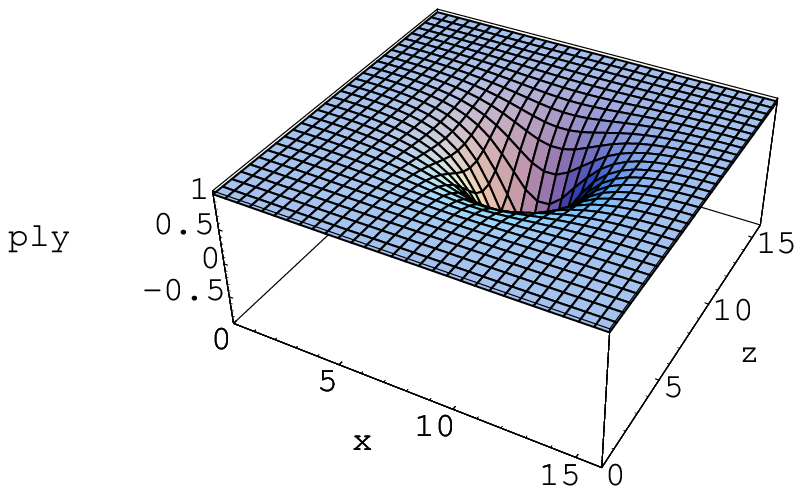}

\caption{Profiles of the action density ($s$), the topological charge density 
($q$), and  of the Polyakov lines ($plt,plx,ply,plz$) as in Fig.
\ref{fig:zerotemp} but for a $16^4$ lattice caloron
obtained from a discretized KvB solution corrected by some cooling to the
one-instanton action plateau (309 cooling steps).
The center of the caloron (at the maximum of its action density) was placed 
to the site $(x,y,z,t)=(8,8,8,1)$. The planes shown in the figures cross 
this point.}
\label{fig:kvb}
\end{figure*}
By that time the boundary artefacts have disappeared (see Fig. \ref{fig:kvb}), 
and the Wilson action has become $S_{inst} = 2 \pi^2 \beta$. The KvB caloron is 
now adjusted to periodic boundary condition also in the $x,y,z$ - directions.
Despite the large separation the constituents formally have, judging according to
the action density it is a non-dissociated caloron. 
It is seen that the cooling does not influence significantly the structure 
of the Polyakov lines of the KvB caloron witnessed by the time-directed Polyakov 
line ($plt$) showing the double peak structure mentioned above.
The space-directed Polyakov lines ($plx, ply, plz$) have a simple structure 
characteristic for trivial holonomy. It is analytically clear that for caloron 
solutions asymptotically $plx, ply, plz \rightarrow \pm 1$.  
This is in contrast to the would-be ''caloron candidate'' obtained by 
cooling from confining equilibrium lattice configurations 
at zero temperature (Fig. \ref{fig:zerotemp}) where the double peak structure is 
present for all $t,x,y,z$ - directions and, generically, the asymptotic holonomy 
in all directions is non-trivial.

\section{Conclusions and perspectives}
\label{sec:conclusions}
In the present investigation we have subjected equilibrium lattice gauge fields
corresponding to various temperatures to ordinary relaxation (usually called 
''cooling'') in order to obtain an ensemble of classical solutions for further study.
In this way we have extracted lowest-action classical 
solutions of unit topological charge typical for the given equilibrium ensemble.
Notice that, at least as long this technique is used, this possibility is restricted 
to the confinement phase. We do not claim to find the real and complete topological 
structure hidden under the quantum fluctuations and consisting of topological lumps 
of compensating sign. What we wanted to see were the simplest solutions suitable as 
building blocks for a semi-classical modelling of the Yang-Mills path 
integral. In conclusion we can say, that cooling of equilibrium lattice fields in 
$SU(2)$ lattice gauge theory shows that there exist topological objects with a
dyonic substructure, that we are able to resolve only in the confined phase. With 
increasing temporal lattice extent it becomes more probable
that the observed dyons recombine into calorons
such that it becomes impossible to perceive the substructure 
looking {\it exclusively} at the distribution of action and topological charge.
Strictly speaking, we have found that the fraction of dissociated calorons
among all single-calorons events does not depend on the physical temperature 
of the equilibrium fields but on the geometric aspect ratio of the box, 
{\it i.e.} the asymmetry of the lattice.  
However, all calorons have a nontrivial holonomy which is 
mapped out by the behavior of the Polyakov line {\it inside and outside} these 
configurations. 

In the limiting case of zero temperature ({\it i.e.} on a symmetric lattice) 
topological lumps with $Q=\pm 1$ obtained by cooling look instanton-like and, 
at the same time, have the characteristic double peak structure of the Polyakov 
line in {\it all} $t,x,y,z$ directions. This distinguishes them from the real
KvB solutions which possess this structure only with respect to a distinguished 
''time'' direction.

Strictly speaking, an analytic solution of the Euclidean equations
of motion with $Q=\pm1$ is impossible on the $4D$ torus \cite{BraamBaal}. 
Nevertheless, on the lattice quasistable solutions of this kind exist. 
If considered only as lumps of action and topological charge, there is no 
contradiction with the previous observations from ''instanton searches''. 
As our analysis shows, for temperatures much lower than the deconfinement 
temperature rotationally symmetric (in $4D$) and (anti)selfdual lumps seem 
to be preferred under cooling. 

At finite temperature, they can be subsumed under the general class of KvB
solutions. For zero temperature, however, a so far unknown parametrization 
with non-trivial asymptotic holonomy has yet to be found.

After having completed the present investigation we were informed by
Ch. Gattringer and R. Pullirsch~\cite{GATTR-PULLIRSCH} about their paper 
''Topological lumps and Dirac zero modes in $SU(3)$ lattice gauge theory on 
the torus'' prior to publication. 
In this paper the authors concentrate on and more systematically continue the
inspection of the low-lying modes of the chirally improved Dirac operator 
on the 4-torus, in the background of equilibrium lattice gauge fields at $T=0$. 
It is extremely interesting that they find, for a certain fraction of Monte Carlo
configurations in the $|Q|=1$ sector, a similar pattern of hopping zero-modes as 
a function of varying fermionic boundary conditions as for $T \ne 0$. 
Moreover, the change of localization happens independently of which of the four 
directions is chosen as ''imaginary time'' direction along which periodicity can be 
purposedly changed. 
The concurrent zero mode positions (two or three as in the finite temperature case) 
are consistent with being randomly distributed in the $4D$ periodic box. The authors 
argue that this observation hints at the existence of a semiclassical background 
consisting of localized (in $4D$) instanton constituents. Whereas other (monopole ?) 
properties of the constituents are less obvious, a non-integer topological charge 
of the constituents has been hypothetically assumed in analogy with the dissociated 
KvB solutions. In fact, this should not be too difficult to be established.
It is in particular this last interpretation that has to pass further tests. 
In case it becomes confirmed then it would be difficult to reconcile this with our 
observations based on cooling. Our findings are consistent with a picture in which 
(at finite temperature) the temporal size of the box determines (inversely) the 
size of the background solutions. In contrast, the scenario of 
Ref.~\cite{GATTR-PULLIRSCH} seems to imply a complete dissolution (within the 
available 4-volume) of some constituents which still span a coherent semiclassical 
background. This would mean that there is no scale of coherence which is dynamically 
generated and decoupled from the overall size of the box. If the latter picture can 
be confirmed we would have to blame cooling (or the cooling action we have used) 
for artificially driving all semiclassical configurations into integer-charged 
topological lumps at lower temperatures whereas separated monopole constituents 
are correctly reproduced by cooling only at higher temperatures. 
This would explain why previous studies 
of the topological properties using various cooling (smoothing) techniques (which 
actually missed the relevant temperature range) were not suitable to discover the 
non-trivial holonomy accompanying all topological charges.

There is still an ongoing debate on the question in as far a semi-classical
interpretation of the QCD vacuum is valid at all \cite{Horvath_etal}.
For the time being we have nothing to add to this discussion.
But our results point to the fact that the instanton gas or liquid model
(as well as other models) have to be reconsidered by taking into account 
the non-trivial holonomy structure. We believe that this will improve the bad
performance of (trivial holonomy) instantons in comparison with lattice smoothing 
results as reported in \cite{Kovacs,DeGAH}.
   
\section*{Acknowledgements}
Three of us (E.-M. I., B.V. M. and M. M.-P.) gratefully acknowledge the kind 
hospitality extended to them at the Instituut-Lorentz of the Universiteit Leiden,
where this paper became finalized. We thank Pierre van Baal, Falk Bruckmann,
and Christof Gattringer for useful discussions and e-mail
correspondence in the final stage of the work.
This work was partly supported by RFBR grants 02-02-17308, 
03-02-19491 and 04-02-16079, 
DFG grant 436 RUS 113/739/0 and RFBR-DFG grant 03-02-04016,
by Federal Program of the Russian Ministry of Industry,
Science and Technology No 40.052.1.1.1112. Two of us (B.V.M. 
and A.I.V.) gratefully appreciate the support of Humboldt-University Berlin 
where this work was initiated and carried 
out to a large extent. The work of E.-M. I. at Humboldt-University 
is supported by DFG (FOR 465).

\end{document}